\documentstyle[aps,amssymb,preprint,12pt]{revtex}
\input{epsf}
\newcounter{aaa}
\newcommand{\ssy}[5]{#1, {#2} {\bf #3}, #5 (19#4).}
\newcommand{\Ex}[2]{\vysk{Example: #1}{#2}}
\newcommand{\ogr}[2]{#1\,\vrule \,{}_{\displaystyle{}_{#2}}}
\newcommand{\vysk}[2]{\begin{trivlist}\labelsep=0pt\item[\bf #1.]
\ #2\end{trivlist}}
\newcommand{\kart}[4]{\begin{figure}[#1]\begin{center}\leavevmode%
\epsfysize=#2\epsffile{#3.ps}\end{center}\caption{#4}\end{figure}}
\newenvironment{conds}[1]{\begin{list}{(\theaaa)\hfill}{\usecounter{aaa} 
\labelwidth=4em\leftmargin=\labelwidth\labelsep=0pt}
\renewcommand{\theaaa}{#1$_\arabic{aaa}$}}{\end{list}}
\newenvironment{teor}[1]{\begin{trivlist}\refstepcounter{#1}%
\setcounter{aaa}{\value{#1}}\labelsep=0pt\item[\bf #1 \theaaa. ]}%
{\end{trivlist}}
\newenvironment{proof}{\par\noindent\emph{Proof.}}{\par\nopagebreak
\hfill$\square$\par}
\author{S. Krasnikov\thanks{Electronic address: redish@pulkovo.spb.su}}
\title{Quantum field theory and time machines}
\begin{document}
\draft\tighten
\maketitle
\begin{abstract}
We analyze the ``F-locality condition" (proposed by Kay to be a
mathematical implementation of a philosophical bias related to the
equivalence principle, we call it the ``GH-equivalence principle"), which
is often used to build a generalization of quantum field theory to
non-globally hyperbolic spacetimes. In particular we argue that the
theorem proved by Kay, Radzikowski, and Wald to the effect that time
machines with compactly generated Cauchy horizons are incompatible
with the F-locality condition actually does not support the
``chronology protection conjecture", but rather testifies that the
F-locality condition must be modified or abandoned. We also show that
this condition imposes a severe restriction on the geometry of the
world (it is just this restriction that comes into conflict with the
existence of a time machine), which does not follow from the
above mentioned philosophical bias. So, one need not sacrifice the
GH-equivalence principle to ``emend" the F-locality condition. As an
example we consider a particular modification, the ``MF-locality
condition". The theory obtained by replacing the F-locality condition
with the MF-locality condition possesses a few attractive features.
One of them is that it is consistent with both locality 
and the existence of time machines.
\end{abstract}  
\section{Introduction}

In recent years much progress has been achieved toward the development of a
rigorous and meaningful quantum field theory in curved
background (semiclassical gravity). In particular, in the framework
of the ``algebraic approach" (see \cite{KRW} and references there)
for globally hyperbolic 
spacetimes  a complete and self-consistent description was constructed of
the real scalar field obeying the Klein-Gordon equation
	\begin{equation}
	(\square-m^2)\phi = 0.
	\label{Kl}
	\end{equation}
However, there are non-globally hyperbolic spacetimes [e.~g.,\ the
Kerr black hole or spacetimes with a conical singularity (those are
universes containing a cosmic string)] quantum effects in which are of
obvious interest. So it would be desirable to have a theory
applicable to such spacetimes as well.
Unfortunately, global hyperbolicity plays a crucial role in the above mentioned
theory, which therefore cannot be straightforwardly 
extended to the general case.
The desired generalization has not been
constructed so far, but a few ``reasonable candidates for minimal necessary
conditions" \cite{KRW} were considered, that is ``statements
which begin 
with the phrase `Whatever else a quantum field theory (on a given 
non-globally hyperbolic spacetime) consists of, it should at least involve
$\dots$'$\,$" \cite{KRW}. The best-studied candidate necessary
condition is the ``F-locality condition" proposed by Kay \cite{Kay}.
Its importance is in  that it turns out to be
quite restrictive. In particular, a theorem 
was recently proved by Kay, Radzikowski, and Wald , which says, roughly speaking, that the F-locality
condition cannot hold in 
a spacetime containing a time machine with the compactly generated Cauchy
horizon \cite{KRW}.
\par The present paper is devoted to the problem of how the
F-locality condition can be amended. The necessity of the amendments
[revealed by the  Kay-Radzikowski-Wald (KRW) theorem] stems from the fact that
	\begin{quote}
	One cannot \emph{just forbid} time machines!
	\end{quote}
It is about six years now that a mechanism which could ``protect
causality" \cite{Conj} against time machines is actively looked for.
The driving force for this search is apparently the idea that the
existence of a time machine would defy the usual notion of free
will. And this would be the case indeed if we found a paradox (like
that usually called ``the grandfather paradox''). For, suppose we
found such a system and such its initial (that is fixed to the past
of the time machine creation) state that the equations governing its
evolution have no solution due to the nontrivial causal structure of
the spacetime.  We know that the system being prepared in this state must
evolve according just to those equations (to change them we must have
confessed that we overlooked some effects, which would have implied
that we simply built an improper model, but not found a paradox) and
at the same time we know that they have no solution.  So we have to
conclude that such an initial state somehow cannot be realized, that
is ``\dots if there are closed timelike lines to the future of a
given spacelike hypersurface, the set of possible initial data for
classical matter on that hypersurface \dots [is] heavily
constrained compared with the same local interactions were embedded
in a chronology-respecting spacetime" \cite{Deu}. 
The dislike for such a contradiction with ``a simple notion of free
will" \cite{HawEl} was so strong that Rama and Sen in
\cite{RaSe}, Visser in \cite{Viss}, and in fact Hawking and Ellis in
\cite{HawEl} proposed just \emph{to postulate} the impossibility of
time machines. Also a postulate prohibiting time machines is
implicitly contained (as is shown by the KRW theorem) in 
Kay's F-locality condition (from now on
by a ``time machine" we mean exclusively a time machine with
the compactly generated Cauchy horizon). The irony of the situation is
that, while no paradoxes have been found so far \cite{Par}, such 
postulates in the absence of a mechanism that could enforce them lead
to precisely the same constraints on one's will.  Indeed, we
know that there are initial conditions on the metric and the fields
such that, when they are fixed at a spacelike surface\footnote{Actually,
even on a part of the surface $t=0$ of an ``almost Minkowskian" space
(cf.~\cite{Conj}).}, the Einstein equations coupled with the
equations of motion for these fields lead to the formation of a time
machine. So if a postulate forbids time machines we
only can conclude either that
(1) there are some
(e.~g.,\ quantum) effects which we have overlooked and which being
taken into consideration always change the equations of motion so that
the time machine does not form, or that (2) such initial conditions
are somehow forbidden. Both possibilities were considered in the
literature.\par\noindent
(1). A popular idea was that the vacuum polarization near a would-be Cauchy
horizon (when it is compactly generated) becomes so strong that its
back-reaction on the metric prevents the formation of the horizon. This
idea, however, has never been embodied in specific results.
The vacuum polarization in spacetimes with a time machine was evaluated
for a few simplest cases \cite{Conj,Pol,Qua}\footnote{There are also
papers where (for non simply connected 
time machines) different results  based on the
``method of images" are obtained and discussed. 
This method, however, involves manipulations with incurably
ill-defined entities and generally allows one to obtain almost any result one
wants (see \cite{Qua} for a detailed discussion).} and it turned out
that sometimes it diverges on the Cauchy horizon and sometimes it does not 
(in the perfect analogy with, say, the Minkowski
space). So it is unlikely  that this effect could always protect causality.
\par\noindent
(2). It is possible that initial data leading to the formation of a time
machine are forbidden not by a restriction on our will, but simply by
the fact that they require some unrealizable conditions. It was shown
\cite{Conj},
for example, that to create a time machine of a non-cosmological
nature (that is evolving from a non-compact Cauchy surface) one has
to violate the Weak energy condition (WEC) and a number of restrictions were
found on such violations (see, e.~g.,~\cite{WEC}). None of them,
however, has been 
able to rule time machines out. Moreover, recently a classical
scenario for WEC violations was proposed \cite{WECvio}.
\par
Thus causality remains still unprotected and any postulate
prohibiting time machines 
without adducing a mechanism that enforces this
prohibition raises the alternative of rejecting either the postulate,
or the idea that whether one can perform an experiment does not
depend on whether causality still holds somewhere in the future.
\par
In the case of the F-locality condition the alternatives at first
glance seem equally unattractive since this condition is based on the
GH-equivalence principle (see Section~\ref{Fl}). However a closer inspection
shows that the F-locality condition does contain a strong arbitrary
requirement (in Section~\ref{why} we discuss this fundamental point
in great detail). So one can reconcile the GH-equivalence principle with
quantum field theory in spacetimes with a time machine by just
abandoning this requirement.   In doing so one still can use the
GH-equivalence principle in the theory. It is only necessary to find its
another mathematical implementation. As an example we consider in
Section~\ref{MF} the ``MF-locality condition''. An important point is
that while expressing the GH-equivalence principle (and seemingly doing it
more adequately than the F-locality condition), \emph{it does not
forbid time machines.} From this we conclude in particular that,
contrary to what was claimed in \cite{KRW} and in a number of
succeeding papers, the KRW theorem does not at all ``provide strong
evidence in support of Hawking's chronology protection conjecture".
It rather rules out the F-locality condition.
\section{Geometrical preliminaries}

%
%
An important role in our discussion will be played by the notion of
\emph{global hyperbolicity}. Globally hyperbolic (GH) spacetimes  
most adequately meet the concept of a ``good" or ``usual" spacetime
(the Minkowski spacetime, for example, is GH).
\begin{teor}{Definition} A subset $N$ of a spacetime $(M,g)$ is
called \emph{globally hyperbolic} if strong causality holds in $ 
N$ and for any points $p,\, q \in N$ the set $J^+(p) \cap J^-(q)$ is
compact and lies in $N$.
\end{teor}
Whether or not a neighborhood $  N\subset M$ is GH is not determined
exclusively by its geometry.  Due to 
the requirement that $\big(J^+(p) \cap J^-(q)\big)\subset N$ it may happen
that $N$ is not GH even though $({  N},\ogr{g}{  N})$ is GH
when  it is regarded as a spacetime in its own
right. So to describe the geometrical properties of a neighborhood
proper we introduce a new\footnote{Connected IGH neighborhoods
were called \emph{locally causal} in \cite{Yur}.} notion:
\begin{teor}{Definition} We call a subset $  N$ of a spacetime $(M,g)$
\emph{intrinsically globally hyperbolic} if $({  N},\ogr{g}{  N})$ is a
globally hyperbolic (GH) spacetime.
\end{teor}
Clearly, whether a neighborhood $ N$ is an intrinsically GH neighborhood (IGHN) 
 does not depend on
the geometry of $M- N$ (in contrast to whether it is a GHN).
%
%
To avoid confusion, note that our notion of ``global hyperbolicity"
is that of \cite{HawEl} and \emph{differs} from that in
\cite{KRW,Kay}. The latter corresponds to our ``intrinsic global
hyperbolicity".
For later use let us list a few obvious properties of (intrinsically)
globally hyperbolic neighborhoods [(I)GHNs]:
	\begin{conds}{GH}
	\item An intersection of two (I)GHNs is an (I)GHN;\label{inter}
	\item Any GHN is an IGHN and an IGHN is a GHN iff
	it is causally convex
	(that is iff no causal curve leaving the IGHN returns in it).
	\label{conv}\setcounter{table}{\value{aaa}}
	\end{conds}
Thus, intrinsic global hyperbolicity is a weaker condition than
global hyperbolicity. In particular, 
\begin{conds}{GH}
\setcounter{aaa}{\value{table}}
\item For any
point $P\in M$ and any its neighborhood $ V$ there exists an IGHN ${
N}:\:P\in N\subset V$, while such a GHN exists if and only if strong
causality holds in $P$. \label{okr} 
\end{conds} 
Property (\ref{conv})
enables us to construct a simple and useful example of
a connected 
IGH but not a GH subset of the (3-dimensional) Minkowski
space\footnote{The existence of such a neighborhood was mentioned in
\cite{Kay} with reference to Penrose.}. 
\kart{h}{0.4\textwidth}{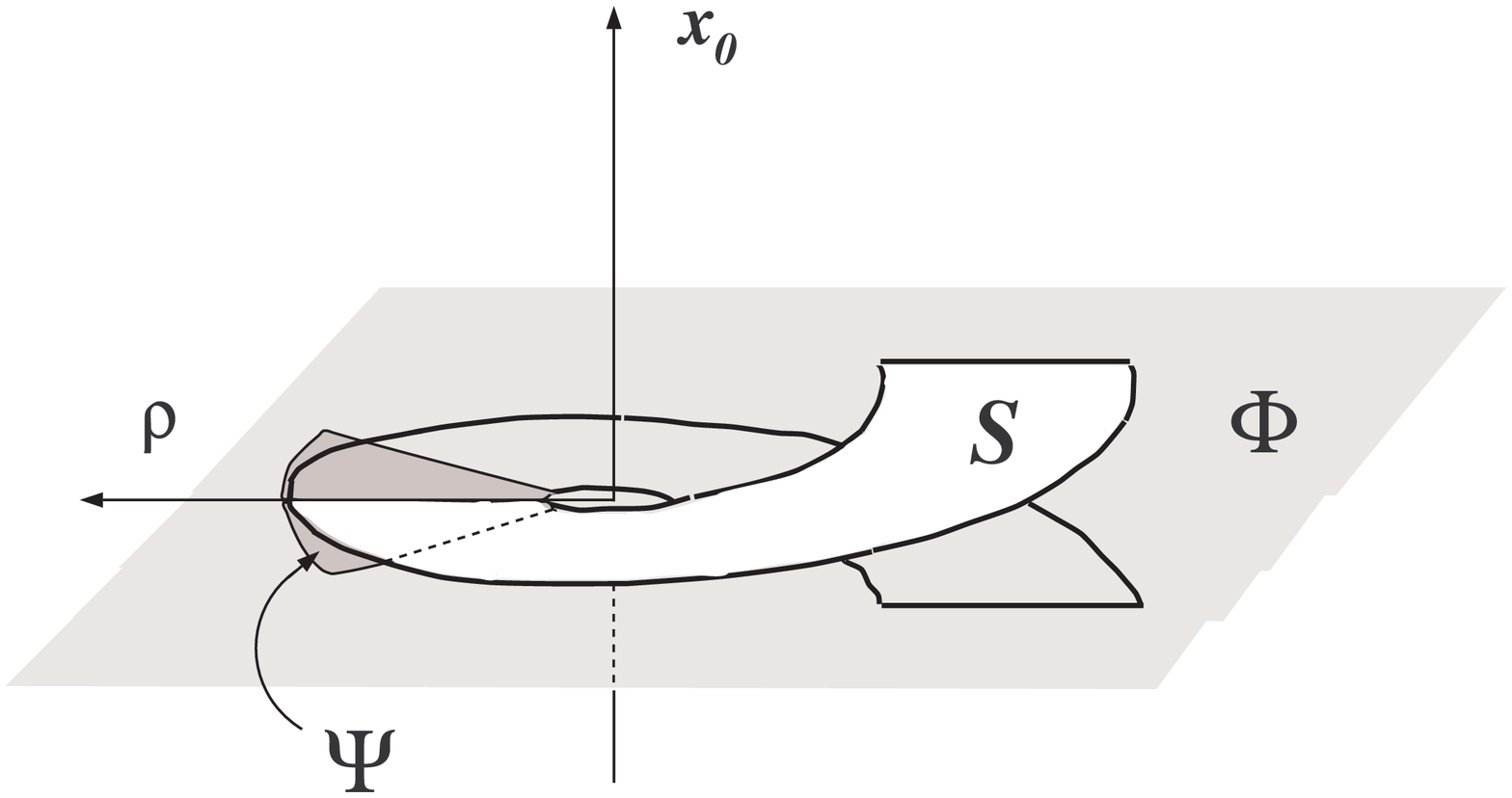}{Construction of a ``bad" set\label{Fig1}}
%
%
\Ex{A ``bad" set}{%
Let ${  V}$ be the cube $\{ x_k \in (-4,4)\}$.
Consider the strip $S\subset   V$ (see Fig.~\ref{Fig1}) given  by the system
	\begin{equation}
	x_0=\varphi/2, \quad \varphi\in[-\pi,\,\pi], 
	\quad \rho \in [1,\,2], 
	\label{strip}
	\end{equation}
where $\rho,\,\varphi$ are the polar coordinates on the plane
$x_1,\,x_2$. 
There are causally connected points on $S$
and, in particular, there are points connected by null geodesics
lying in $V$
(or \emph{null related in $V$}, in terms of \cite{KRW,Kay}).
A simple calculation based on the fact that
	\begin{equation}
	{\mathbf A }\mbox{ is spacelike}
	\quad \mbox{whenever} \quad | A_0/A_i| <1 
	\label{nakl} 
	\end{equation}
shows, however, that
	$$
	v_1,\, v_2 \in S, \quad  v_1 \ne v_2, \quad
	 v_1 \preccurlyeq v_2 \qquad \Rightarrow \qquad
	\varphi(v_2) - \varphi(v_1) > \varphi_0 > \pi.
	$$
So a causal curve can connect two points in $S$ only if
one of them lies above the plane $\Phi \equiv\{v| \, x_0(v)=0\}$ and
the other below $\Phi$. Hence,
	\begin{enumerate}				
	\item[(\emph{a}) ] All causal curves connecting
	points of $S$ intersect 
	the plane $\Phi$.
	\end{enumerate}
Similarly, by  simple though tiresome considerations one can show that
	\begin{enumerate}				
	\item [(\emph{b}) ] There is a closed set  $\Theta \subset \Phi$ 
	such that $S \cap \Theta = \emptyset$ 
	and none of the causal curves from $S$ to $S$ 
	intersects $\Psi\equiv\Phi -\Theta$. 
	\end{enumerate}
(For example, we can choose $\Psi\equiv  \{v\in \Phi|\:
\rho(v) \in (0.8,\,2.2),\: |\varphi(v)| < 0.1\}$.)
Consider now $S$ as a subset of the spacetime $M' \equiv M - \Theta$.
Properties (\emph{a,b}) ensure that $S$ is a (closed) achronal set
and hence by Prop.~6.6.3 of \cite{HawEl} the interior $B$ of its
Cauchy domain in $M'$ is a GH subset of $M'$. Thus by (\ref{conv})
$B$ is an IGHN and not a GHN.}

Note that we have used the fact that $M$ is the Minkowski space
only in stating
(\ref{nakl}). It can be easily seen, however, that within \emph{any}
neighborhood in \emph{any} spacetime coordinates $x_i$ can be found 
such that (\ref{nakl}) holds in the cube $\{ x_k \in (-4,4)\}$. So (being
generalized to the 4-dimensional case) this example proves 
the following proposition:
%
%
\begin{teor}{Proposition} \it
For any point $p$ and any its neighborhood $  V$ such a connected IGHN
${ B}\subset V $ of $p$ and such a pair of null related in $V$ points $r,\,q
\in B$ exist that $r$ and $q$ are not connected by any causal
curve lying in $ B$. \label{badset}
\end{teor}
%
%
\section{F-locality}
\label{Fl}
The algebraic approach to quantum field
theory (below we cite only some basic points that have to do with
F-locality, for details see \cite{KRW} and references therein) 
is based on the notion of the ``field algebra", which is
a $*$-algebra with identity $I$ generated by
polynomials in ``smeared 
fields'' $\phi(f)$, where $f$ ranges over the space 
${\cal D}( M)$ of
smooth real valued functions compactly supported on $M$.
The  smeared 
fields  $\phi(f)$ are just some abstract objects (informally they can be 
understood as $\phi(f)=\int_M\phi(x)f(x)\sqrt{-g}\,d^4x$, where
$\phi(x)$ is the 
``field at a point" operator of the (non-rigorous) conventional QFT).
A field algebra is defined by the relations
(for all $f, h \in {\cal D}(M)$ and for all pairs of
real numbers $a,b$): 
	\begin{equation}
	\phi(f)=\phi(f)^*, \quad
	\phi(a f + b h)=
	a\phi(f)+b\phi(h),\quad
	\phi((\square-m^2)f)=0
	\label{Pre}
	\end{equation} 
(defining a ``pre-field algebra") and a relation fixing commutators $
[\phi(f),\phi(h)]$, which we discuss in the following subsections.
\par
Given a field algebra one can proceed to build a complete
quantum theory of the free scalar field by introducing the notion of
\emph{states}, postulating some properties for  ``physically
realistic'' states and prescriptions for evaluating physical quantities 
(such as the renormalized
expectation value of the stress-energy tensor) for
these states. We will not go into this ``second level" \cite{Kay} of
the theory.

\subsection{The globally hyperbolic case}

\begin{teor}{Definition}
Let $\cal E$ be a subset of ${\cal D}(M)\times {\cal D}(M)$ and let
$\triangle$ be a functional on pairs $f,\,h$, where $f,\,h \in {\cal
D}(M)$ and $(f,h)\in {\cal E}$. We shall call $\triangle$ 
\emph{a bidistribution on $\cal E$} if it is separately
linear and continuous (with respect to topology of ${\cal D}(M)$) in
either variable. 
\end{teor}
To fix a commutator relation for the field algebra consider the
Klein-Gordon equation 
(\ref{Kl}) given on an IGHN $U$. Let $\triangle$ be its
bidistributional solution, that is, a bidistribution on 
${\cal D}(U)\times {\cal D}(U)$ satisfying $\triangle((\square
-m^2)f,h)=\triangle(f,(\square-m^2)h)=0$ for all $f,h \in
{\cal D}(U)$. Among all such solutions there is a preferred one:
\begin{teor}{Definition}
Let $\triangle^{A(R)}_{ U}$ be the fundamental solutions of
the inhomogeneous Klein-Gordon equation on a neighborhood $U$ satisfying
the property: 
	\begin{equation}
	\triangle^{A(R)}_{ U}(f,g) = 0
	\quad \mbox{whenever} \quad
	\mathop{\rm supp} f \cap 
	J^\mp(\mathop{\rm supp} g, U) = \emptyset .
	\label{LCau}
	\end{equation}
Then we call a bidistributional solution of
the \emph{homogeneous} Klein-Gordon equation $\triangle_{ U} \equiv
\triangle^A_{ U} - \triangle^R_{ U}$ \emph{the advanced minus
retarded solution  on $U$}. 
\end{teor}
It turns out that for any IGHN $U$, $\triangle_{ U}$ exists and is
unique. So we complete the definition of a ``field algebra" by adding
to (\ref{Pre}) the following commutator relation:
	\begin{equation}
	[\phi(f),\phi(h)]=i\triangle_M\,(f,h) I. 
	\label{Com}
	\end{equation}
Which of the bidistributional solutions of equation (\ref{Kl}) is the
advanced minus retarded solution for a given region $U$ is completely
determined by the causal structure of $U$. This allows one to prove
the following important fact \cite{Kay}:
\vysk{The F-locality property (Form I)}{Every point $p$ in a GH spacetime
$M$ has an intrinsically 
globally hyperbolic neighborhood ${   U_p}$ such that for all  $f, h\in
{\cal D}(U_p)$, relation (\ref{Com}) holds with
$\triangle_M$ replaced by $\triangle_{U_p}$.}
We can also reformulate the  F-locality property in a slightly
different form by ``gluing" all these $\triangle_{U_p}$ into a single
bidistribution $\triangle^F$ .\par
%
Let $\triangle $ be  a bidistribution on 
 ${\cal E} \supset {\cal D}(U)\times
{\cal D}(U)$. It induces a  bidistribution  $\ogr{\triangle }{ U}$ on
${\cal D}(U)\times {\cal D}(U)$ by the
rule
	$$
	\forall f,h \in {\cal D}(U) \qquad
	\ogr{\triangle }{   U}(f,h) \equiv \triangle(f,h).
	$$
\begin{teor}{Definition}
We shall call ${   U}$ and $\triangle $ \emph{matching}
if ${   U}$ is a connected IGHN and
	$$
	\ogr{\triangle }{   U} = \triangle_{   U}.
	$$
\end{teor} 
\vysk{The F-locality property (Form II)}{There are such  an
open covering of a GH spacetime $M$ by IGHNs $\{U_\alpha\}$ and such 
 a bidistribution $\triangle^F$ on ${\cal E}_U$ that 
	\begin{conds}{Pr}
	\item $\triangle^F$ matches any $U_\alpha$
	\item When $(f, h) \in {\cal E}_U$,
	relation (\ref{Com}) holds with $\triangle_M$ replaced
	by ${\triangle^F}$.
	\end{conds}
	}
Here and subsequently if $\{U_\alpha\}$ is a set of neighborhoods in $M$ we
write ${\cal E}_U$ for $\bigcup_\alpha [{\cal D}(U_\alpha)\times {\cal
D}(U_\alpha)]$. 
%
\subsection{The non-globally hyperbolic case}
To build a field algebra in a non-globally
hyperbolic spacetime we can start with a ``pre-field algebra"
(\ref{Pre}). Then, however, we meet a problem with the commutator
relation since $\triangle_M$ is (uniquely) defined only for GH
spacetimes and there are no
obviously preferred solutions of (\ref{Kl}) any longer. So, we need
some new postulate and Kay proposed \cite{Kay} to infer such a postulate
from the equivalence principle, which as applied to our situation he
formulated as follows.
\vysk{The GH-equivalence principle}{On an
arbitrary spacetime, the laws in the small  
should coincide with the ``usual laws for quantum field theory on
globally hyperbolic spacetimes".}
%
%
From this principle he \emph{postulated} in a sufficiently small
neighborhood of a point in an arbitrary spacetime what
holds by itself in a GH spacetime. Namely, he requires:
\vysk{The F-locality condition (Form I)}{Every point $p$ in $M$ should have an
intrinsically globally hyperbolic neighborhood ${ U_p}$ such that, for
all $f, h\in 
{\cal D}(U_p)$, relation (\ref{Com}) holds with $\triangle_M$ replaced 
by $\triangle_{   U_p}$.}
It is implied that a spacetime for which there is no field algebra
satisfying this condition (a ``non-F-quantum compatible" spacetime)
cannot arise as an approximate description of a state of quantum
gravity and must thus be considered as unphysical.\par
To reveal the logical structure of the F-locality condition
we reformulate it analogously to the F-locality property.
%
%
\vysk{The F-locality condition (Form II)}{
There should be such  an
open covering of a spacetime $M$ by IGHNs $\{U_\alpha\}$ and
 a bidistribution $\triangle^F$ on 
${\cal E}_U$ that 
	\begin{conds}{Con}
	\item $\triangle^F$ matches any $U_\alpha$\label{Fl1};
	\item When $(f, h) \in {\cal E}_U$,
	relation (\ref{Com}) holds with $\triangle_M$ replaced
	by ${\triangle^F}$. \label{Fl2}
	\end{conds}
	}
\par
An important difference between these two parts of the F-locality
condition is that (\ref{Fl2}) just specifies what algebra we take to
be the ``field algebra", while (\ref{Fl1}) is a nontrivial
requirement placed from the outset upon the spacetime. It is
significant that the proof of the KRW-theorem rests upon (\ref{Fl1}).
\par
The F-locality condition clearly does not fix
\emph{all} commutators. The value of $[\phi(f),\phi(h)]$ remains
still undefined for $f,\,h$ whose supports do not belong to a common
${U_\alpha}$. It is more important, however, to find out whether
this uncertainty extends to arbitrarily small regions. Indeed, 
to find such local quantities as
$\left<T_{\mu\nu}\right>(p)$ or, say, $\left<\phi^2\right>(p)$ it
would be enough 
to know all commutators $[\phi(f),\phi(h)]$ with functions $f,\,h$
both supported in a small neighborhood $V$ of $p$. This leads us to
the following question:
Is it true for at least \emph{some} open covering $\{V_\alpha\}$
that 
	\begin{equation}
	\forall (f,h) \in {\cal E}_V \qquad
	{\triangle}'^F(f,h) = {\triangle}^F(f,h) \label{Rav}
	\end{equation}
whenever both $\triangle^F$ and
${\triangle}'^F$ satisfy (\ref{Fl1}) (with possibly different
$\{U_\alpha\}$)?
It turns out that the answer is negative even in the simplest case.
Indeed, if $M$ is the Minkowski space and
$\triangle^F$ is a  solution of (\ref{Kl}) satisfying
(\ref{Fl1}), then so is ${\triangle}'^F$:
	$$
	{\triangle'}^F(f,g) \equiv \triangle^F(f',g), \qquad
	\mbox{where }f'(x^\mu) \equiv f(x^\mu) + f(x^\mu + a^\mu)
	$$
and where by $a^\mu$ we denote an arbitrary constant spacelike vector field.
Clearly, for any $\{V_\alpha\}$  we can find an  $a^\mu$ such that (\ref{Rav})
breaks down.
\par
So the F-locality condition was proposed only as a \emph{necessary}
condition which is to be supplemented with conditions of ``the second
level" to obtain a complete theory.
%
%
\section{The paradox and its resolution}
\label{why}

The F-locality condition (or (\ref{Fl1}) to be more specific) includes
actually a postulate forbidding time machines. This follows from the
Kay-Radzikowski-Wald theorem:
\vysk{The KRW theorem}{\em If a spacetime has a time machine with
the compactly generated Cauchy horizon, then there is no extension to $M$ of
the usual field algebra on the initial globally hyperbolic region $D$
which satisfies the F-locality condition.}
Here by ``the usual field algebra" an algebra is meant where for 
$f, h \in {\cal D}(D)$ relation (\ref{Com}) holds with $\triangle_M$
replaced by ${\triangle_D}$ (for the proof of the theorem and the
precise definition of $D$  see \cite{KRW}.)
\par
As is discussed in the Introduction, postulating causality without
adducing a ``protecting" mechanism, one comes up against a
contradiction with the usual notion of free will, which can be regarded as
a paradox.
\par
Such a situation (when a paradox arises from postulating in the general
case a condition harmless in the GH case) is in no way strange or new. 
\Ex{Classical pointlike particles}{%
Consider a system of elastic classical balls.
As long as one studies only GH spacetimes one sees that the following
property holds\footnote{A
model describing such a system can be found in \cite{Par}. A
specific mathematical meaning is assigned there to the words ``a
world line of a ball", etc. The property then can be \emph{proven} within
this model.}
\vysk{\textmd{\sl ``The property of balls conservation"}}{Any Cauchy
surface intersects the same number of the world lines of the balls.}
Going to arbitrary spacetimes, one finds that the evolution of a system of
balls is no  longer uniquely fixed by what fixes it in the GH case.
 To overcome this problem (in the perfect
analogy with the F-locality condition) one could 
adopt the following postulate\footnote{Such an approach
was really developed in a number of works (e.~g.\ see 
\cite{Deu,RaSe,Cons}).} (note that in the general case it is just a
postulate, that is an \emph{extraneous (global) condition}
and not a consequence of any other local principles accepted in the model):
\vysk{\textmd{\sl``The condition of balls conservation"}}{Any partial Cauchy
surface should intersect the same number of the world lines of the
balls.} Then one would find \cite{RaSe,Par} that there are
``non-classical compatible" 
spacetimes (e.~g.,\ the Deutsch-Politzer space) that are spacetimes in
which initial data (i.~e., data at some partial Cauchy surface) exist
incompatible with the ``Postulate of balls conservation".
This fact constitutes an
(\emph{apparent}, see \cite{Par}) paradox and so one could claim
that the existence of such paradoxes suggests that time machines are
prohibited \cite{RaSe}. On the other hand, as we discussed above, it
seems more reasonable to look for contradictions 
which we ourselves could introduce in the model in the process of
constructing. In doing so we would interpret the ``non-classical
compatibility" of the Deutsch-Politzer spacetime  as  evidence
not against the realizability of this spacetime, but rather against the
postulate. Indeed, abandoning this postulate we resolve
the paradox (and thus permit time machines) while causing no harm to any
known physics \cite{Par}.
}
The above example suggests that to avoid the
difficulties connected with forbidding the time machine, which we
discussed in the Introduction, it would be natural just to abandon the
F-locality condition. The problem, however, is that while we can
easily abandon ``the postulate of balls conservation", the F-locality
condition seems to be based on the philosophical bias resembling the
equivalence principle, which is something one would not like to
reject. So, in the remainder of the Section we show that  the F-locality
condition contains actually an \emph{arbitrary \emph{(i.~e., not implied by
the GH-equivalence principle or any other respectable physical
principle)} global requirement} and therefore can be rejected
or modified without regret.

%
\begin{teor}{Proposition} \it For any $\triangle $ and any neighborhood ${ 
V}$ there exists a connected IGHN ${  B} \subset {  V}$ that does
not match $\triangle $.
\end{teor}
\begin{proof} Without loss of generality (see (\ref{okr})) ${  V}$ may be
thought of as being an IGHN. So either ${  V}$ itself is the desired
neighborhood or $\ogr{\triangle }{  V}$ is the advanced minus
retarded  solution $\triangle_{  V}$ on $  V$. In the latter case we can
simply adapt the proof of the KRW theorem \cite{KRW} for our needs.
Namely, let ${  B}$ be the set 
from Prop.~\ref{badset} and $r,\,q$ the points appearing there. To obtain a
contradiction suppose that ${  B}$  matches
${\triangle }$ and hence matches $\ogr{\triangle }{ V}=\triangle_{
V}$ also. This would mean, by definition, that
	\begin{equation}
	\ogr{(\triangle_{  V})}{  B} = \triangle_{  B},
	\end{equation}
but $\triangle_{  B}(r,q)=0$ since $r$ and $q$ are not causally
connected in ${  B}$, while $\ogr{(\triangle_{  V})}{  B}$ 
is singular
at the pair $r,\,q$ (see \cite{KRW} for the proof) since both of these
points belong to ${ V}$ and are null connected in it. Contradiction. 
\end{proof}
Thus we see that even if a spacetime is globally hyperbolic there are two
families of IGHNs for any its point: causally convex (and thus GH) sets 
$\{{ G}_\alpha\}$ (let us call them \emph{``good"}) and those
containing null related points that are intrinsically non-causally connected
(we shall call them \emph{``bad"} and denote by $\{{ B}_\beta\}$). Both
families include 
``arbitrarily small" sets (i.~e., for any neighborhood ${  V}$
one can find both a ``good" (${  G}_{\alpha_0}$) and a ``bad" 
(${  B}_{\beta_0}$) subsets of ${  V}$). 
Irrespective of what meaning one assigns to 
the term ``the laws in the small", it seems reasonable to
assume that they are 
the same for ${  B}_{\beta_0}$ and ${  G}_{\alpha_0}$. The
more it is so as an observer cannot determine (by geometrical means) whether
a neighborhood is ``good" or ``bad" without leaving it. We have seen
that the ``good" sets match the commutator function, while
the ``bad" ones do not.
So it follows that the identity of physics in two sets does not imply
that they both 
match the same bidistribution. Correspondingly, the fact that the
laws in a small region coincide with any other laws does
not imply that it (or any its subset) matches the commutator function
on a bigger region. So \emph{the requirement (\ref{Fl1}) that a point should
have a neighborhood matching a global
commutator function
is not an expression of the GH-equivalence principle}, but is
rather an extraneous condition. 
It is also an essentially \emph{global} condition.  Indeed,
for any point one \emph{always} can find a bidistribution matching
\emph{some} IGHN of the point and so the main idea of
(\ref{Fl1}) is that such a bidistribution should exist globally.
\par
We see thus that indeed the F-locality condition needs emendations, since
while leading to possible paradoxes it contains a strong
nonjustified requirement. 

%
%
\section{Modified F-locality}

In this Section we formulate and discuss a candidate necessary
condition alternative to the F-locality condition. Being an implementation of the GH-equivalence principle  (coupled with the locality principle, see below), it nevertheless does not forbid any
causal structure whatsoever. Thus a theory based on this condition is
free from the paradoxes discussed above, which provides  further
evidence in favor of the idea that the existence of time machines is
inconsistent not with  the equivalence principle, but only with its
inadequate implementation. 
\par
Consider a commutator $[\phi(x),\phi(y)]$. Physically this commutator describes the process in which a particle created from vacuum in $x$ annihilates in $y$. So when we require [as we did in (\ref{Com})] that the commutator function should vanish for non-causally connected $x$ and $y$ we just implement the (most fundamental) idea that an event can affect only those events that are connected with it by causal curves or, in other words, that particles (or information in any other form) cannot propagate faster than light. The very same idea (called \emph{locality}, or \emph{causality}, or \emph{local causality} depending on the formulation and application) suggests that if the conditions are fixed in $J^+(x) \cap J^-(y)$ (that is, in all points where a non-tachionic particle propagating from  $x$ to $y$ can find itself), then  $[\phi(x),\phi(y)]$ is thereby also fixed. Thus, from locality it seems natural to require as a necessary condition that the field algebra in a globally hyperbolic neighborhood  $G$ does not `feel' whether or not there is something outside  $G$ [recall that for any $x,\,y \in G$ any point $z \in M- G$ lies off  $J^+(x) \cap J^-(y)$].
We can then construct a field algebra (at least on a part of $M$, see below) by
adopting the following modification of  the F-locality condition:
\vysk{The MF-locality condition}{If $\{{G}_\alpha\}$ is the collection of all globally hyperbolic subsets of a spacetime $M$, then for all 
 $(f, h)\in {\cal E}_{ G}$ relation (\ref{Com})
should hold with $\triangle_M$ replaced by $\triangle^{MF}$ defined
to be a bidistribution on ${\cal E}_{ G}$ matching each ${ G}_\alpha$.} 
\label{MF}
(In other words, we require that $\phi(f)$ and $\phi(h)$ with $f$ and
$h$ supported on a common GHN ${ G}_\alpha$ should commute as if there were no
ambient space $M - { G}_\alpha$ at all.)
This condition obviously holds in a GH spacetime, where\footnote{Generally $\triangle^{MF}$ is not the same as 
$\triangle^F$. This follows directly from the non-uniqueness of $\triangle^F$ shown at the end of Sect.~\ref{Fl}.} 
$\triangle^{MF}=\triangle_{M}$.
\par
\emph{Remark.} In discussing the field algebra we operate with such `non-local' (by their very nature) entities as  commutators $[\phi(x),\phi(y)]$, where $x$ and $y$ can be wide apart. No wonder that relevant statements are also formulated in `non-local' terms. In particular, both the MF-locality and the 
F-locality conditions distinguish some classes of IGHNs of a point from the others. In the former case those are the causally convex neighborhoods and in the latter case the distinguished class is not specified, but its existence is postulated. But to learn whether or not a given IGHN belongs to the distinguished class we have to consider how it is embedded in the ambient space and to take into account properties of this space [e.~g.\ to check whether or not a set $V$ is causally convex one must consider the whole $J^+(V)$]. In this connection we emphasize that the MF-locality condition is  not a non-local postulate  (much less a postulate contradicting locality). That is, it does not require that a spacetime, or a field algebra, possess any non-local properties. On the contrary, we found out what \emph{locality} requires in a specific situation (it is the description of this situation that necessitates non-local terms as we argued above) and chose these requirements as a postulate of the theory\footnote{Note that the same situation takes place in the globally hyperbolic case. The postulate (\ref{Com}) also may seem non-local since the condition defining $\triangle_M$ contains [see (\ref{LCau})] a `non- local' part $\mathop{\rm supp} f \cap 
J^\mp(\mathop{\rm supp} g, U) = \emptyset$.}. 
\par
The MF-locality condition differs from the F-locality condition in 
that
	\begin{itemize}
	\item[\bf A.] \emph{Some IGHN} are replaced by \emph{each GHN} and
	\item[\bf B.] A condition is imposed only on 
	the field algebra, but not on the geometry of the background 
	spacetime.
	\end{itemize}
Correspondingly, two important consequences take place:
\par\noindent {\bf A.} As we discussed in Section~\ref{Fl}
the  F-locality condition does not uniquely fix the commutator function.
Neither does the  MF-locality condition. 
The situation has improved, however, in that now we can fix at least
the ultraviolet behavior of the commutator function in the region
 $G \equiv \bigcup_{\alpha} {  G}_\alpha$ where strong causality holds.

%
\begin{teor}{Proposition} \it
	For any spacetime $M$ with non-empty $G$,
	$\triangle^{MF}$ exists and is unique.
	\label{Uni}
	\end{teor}
\begin{proof} Consider two GHNs ${  G}_i$ and 
${  G}_{k}$. By (\ref{inter}) their intersection is also a
GHN (we can thus denote it ${  G}_j$) and points in ${  G}_j$
are causally related if and only if they 
are causally related in ${  G}_{i}$ (and thus also in ${ 
G}_{k}$). So (see \cite{Kay})
	$$
	\ogr{\left(\triangle_{{  G}_{i}}\right)}
	{{  G}_j}=
	\triangle_{{  G}_j}	=
	\ogr{\left(\triangle_{{  G}_{k}}\right)}
	{{  G}_j},
	$$
which means that we can \emph{define}  $\triangle^{MF}$ by
the equation 
	\begin{equation}
	\ogr{\triangle^{MF}}{{  G}_\alpha} \equiv 
	\triangle_{{  G}_\alpha }.
	\label{locel}
	\end{equation}
This guarantees that $\triangle^{MF}$ is a desired bidistribution
matching all ${ G}_\alpha$. At the same time, \emph{any} functional
matching them must satisfy (\ref{locel}), which proves the uniqueness.
\end{proof}
{\bf B.}
As we discussed above, the F-locality condition is global in
nature. Either $\triangle^F$ does not exist on $M$ and we must
exclude the whole $M$ from consideration or $\triangle^F$ exists and
then no region is distinguished in this sense.  The situation differs
greatly if we postulate the MF-locality condition instead. On the one hand,
\emph{any} spacetime is allowed now
(since $\triangle^{MF}$  always exists
(see Prop.~\ref{Uni}); there are no ``non-MF-compatible spaces") and,
on the other hand, different parts of a spacetime now have different
status.  Namely, each point $p \in G$
has a neighborhood ${  U}_p$ such that $\triangle^{MF}\,(f,h)$ is
determined by the MF-locality condition at least for $f, h\in {\cal
D}_{U_p}$.  
So one can develop the theory as we mentioned in the
beginning of Section~\ref{Fl} 
and  eventually find $\left<T_{\mu\nu}\right>(p)$. But this
cannot be done at this stage for a point in $(M - G)$, where no field
algebra is fixed. Thus 
the surface $\partial G$ separates the area of the present
version of semiclassical gravity from \emph{terra
incognita}\footnote{In this respect $\partial  G$ is similar to
Visser's ``reliability boundary'' \cite{Viss}. The main difference is
that the latter conceptually bounds the region where semiclassical
gravity breaks down because of quantum gravity corrections.}.
\par
The role of $\partial G$ is especially important in the time machine
theory since it is $\partial G$ where the divergence of the
stress-energy is expected by many authors.  So it should be stressed
that \emph{physically} there is nothing particular in points of
$\partial G$ (including the ``base points"; see \cite{KRW}). In 
perfect analogy with coordinate singularities in general relativity,
$\partial G$ does not correspond to any physical entity
and the fact that we cannot find the
energy density in a point of $\partial G$ means not that it is singular or
ill-defined here but simply that we do not know how to do this.
\par
\emph{Remark.} The MF-locality condition was proposed in this paper
primarily to clarify the relation between causality violations and the
GH-equivalence principle. However, the uniqueness proved above and the
simplicity of the underlying physical assumption suggest that perhaps it
deserves a more serious consideration as a possible basis for
constructing semiclassical gravity in non-globally hyperbolic
spacetimes. Then it would be interesting to find out
whether the theory proposed by Yurtsever \cite{Yur} (which does not,
at least explicitly, appeal to any locality principle) is consistent
with it.

\end{document}